# FEW-MODE FIBER TRUE TIME DELAY LINES

**Sergi Garcia, Ruben Guillem and Ivana Gasulla, *Senior Member, IEEE***
*ITEAM Research Institute, Universitat Politècnica de València, 46022 Valencia, Spain*
*e-mail: sergarc3@iteam.upv.es*

**ABSTRACT**
Space-division multiplexing optical fibers can provide not only parallel channel transmission but also parallel distributed signal processing, being this a feature particularly attractive in Microwave Photonics applications. We present here few-mode fiber links with tailored modal propagation and dispersion properties that operate as sampled true time delay lines for radiofrequency signals.

## 1. INTRODUCTION

Space-division Multiplexing (SDM) technologies in optical fibers [1] can be exploited beyond data signal distribution in the framework of high-capacity digital communications systems. We have recently proposed and demonstrated their potential as a compact medium to offer what we coined as "fiber-distributed signal processing", that is, the simultaneous implementation of both signal distribution and processing tasks while the signal is propagated through the optical fiber, [2]. This capability is particularly relevant for next-generation fiber-wireless communications systems, such as 5G radio access networks and the Internet of Things. In this sense, we foresee that microwave photonics (MWP) signal processing and radio-over-fiber distribution [3] can benefit greatly from the use of different SDM fiber technologies in terms of compactness and weight, while assuring broadband versatility, reconfigurability and performance stability. This brings a considerable challenge that implies the development of novel multicore fiber (MCF) [4,5] or few-mode fiber (FMF) [6] solutions where the different spatial paths (cores or modes) translate into the different propagation characteristics (in terms of group delay and chromatic dispersion) required for signal processing. In particular, these SDM solutions are designed to act as optical sampled true time delay lines (TTDLs), which are the basis of most MWP signal processing functionalities, such as signal filtering, optical beamforming for phased-array antennas and arbitrary waveform generation [3]. The goal of a TTDL is to provide a set of frequency independent and tunable delays within a given frequency range.

In the particular case of FMFs, the fiber must be designed so that every mode (or group of modes) experiences, at a given optical wavelength, the adequate group delay to fulfil the main principle for TTDL operation. As Fig. 1 shows, we must obtain at the output of the FMF link, a set of time-delayed replicas of the modulated signal that feature a constant differential delay between them (named as basic differential delay, Δτ), [3]. If only one optical wavelength is implicated, as shown in Fig. 1, the TTDL features 1D performance where all the samples result from exploiting the fiber spatial diversity. But this approach offers in addition 2D operation if we combine the spatial diversity with the optical wavelength diversity provided by the use of multiple optical wavelengths.

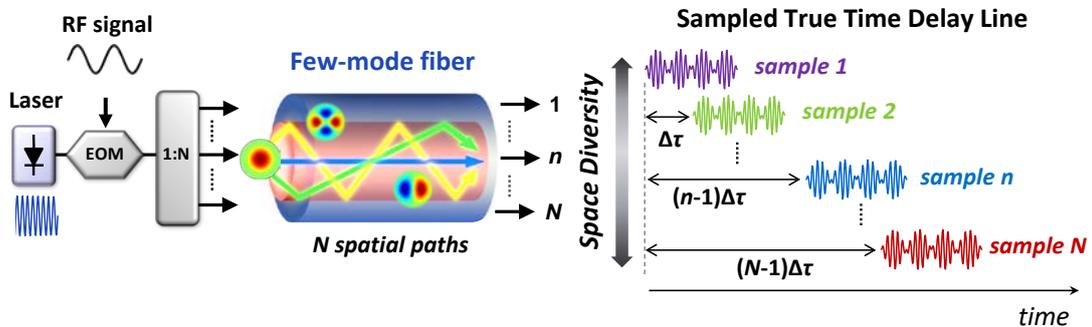

*Figure 1. Tuneable sampled true time delay line based on a few-mode fiber link*

We must assure, in addition, a low level of coupling between groups of modes, what calls for the predilection of refractive step-index (SI) profiles and the use of short distances combined with direct detection techniques. In [6], we reported 3-sampled TTDL operation on a 60-m SI 4-LP-mode fibre where 3 modes were injected at the fibre input and a long period grating (LPG) was inscribed to excite the $LP_{02}$ mode while adjusting the time delay of the sample associated to that mode. Although that scheme resulted in a constant differential delay between samples, it did not allow TTDL tuneability with the optical wavelength.

We propose here a FMF-based 4-sample TTDL approach where we obtain, for the first time to our knowledge, time delay tuneability over a specific optical wavelength range. This approach combines the custom design of a 7-LP-mode ring-core fiber and the inscription of 5 LPGs at the appropriate locations along the FMF to achieve both constant differential time delay and differential chromatic dispersion values between the different signal samples.

## 2. CONCEPT UNDERLYING TUNEABLE TRUE TIME DELAY LINE OPERATION ON FEW-MODE FIBERS

Sampled TTDL operation requires constant basic differential delays $\Delta\tau$ between adjacent samples. In addition, time delay tunability can be added by assuring a linear increment of the basic differential delay with the optical wavelength. In the particular case of a FMF-based TTDL, the different samples can be created in the spatial diversity domain by the use of the different modes that propagate along the fiber. The group delay per unit length of a given LP$_{lm}$ mode, $\tau_{lm}$, can be expanded in 1$^{st}$-order Taylor series around a central wavelength $\lambda_0$ as

$$\tau_{lm}(\lambda) = \tau_{lm}(\lambda_0) + (\lambda - \lambda_0) D_{lm}, \tag{1}$$

where $\tau_{lm}(\lambda_0)$ is the group delay per unit length at $\lambda_0$ and $D_{lm}$ is the chromatic dispersion at $\lambda_0$ for the LP$_{lm}$ mode. For TTDL operability, both $\tau_{lm}$ and $D_{lm}$ must increase linearly between consecutive samples. Thus, a TTDL built upon a single FMF would require the use of a particular set of propagating modes that satisfies these two characteristics. However, this is, in general, impossible to accomplish in FMFs with classic refractive index profiles (i.e., step index, graded index, ring-core index, etc.). And, even in the hypothetical case that it was possible, the differential group delays between modes would probably be extremely high for MWP applications (i.e., very low operating radio frequencies), limiting the fiber length to a few meters. These differential group delays would be in addition much greater than the incremental dispersion effect itself, so that the differential delay tunability would be insignificant.

We propose then to obtain the signal samples as a combination of different modes. A set of LPGs inscribed at specific positions along the FMF link will not only act as mode converters, [7], but will allow as well to control the final group delay and chromatic dispersion values associated to each signal sample. Let's explain this approach considering a given signal sample $i$. This sample will be excited by a given incoming fiber mode that is injected into the FMF. Then, that signal will propagate a certain distance given by $L - l_{lm}^{(i)}L$ (with a certain group delay and chromatic dispersion), where $L$ is the total length of the FMF link, until a given LPG will transform the incoming fiber mode into a different outgoing LP$_{lm}$ mode (with a different group delay and chromatic dispersion). The parameter $l_{lm}^{(i)}$ is defined as the normalized length along which the $i^{th}$ sample is propagated by the outgoing LP$_{lm}$ mode. This process can be repeated as many times as necessary by concatenating different LPGs inscribed at the longitudinal positions $L - l_{lm}^{(i)}L$, where the subindex $lm$ stands for the outgoing mode in each LPG. This way, at the output of the FMF link, the group delay of the $i^{th}$ TTDL sample, $\tau_i$, at a given wavelength $\lambda_0$, can be expressed as a combination of the LP$_{lm}$ modes involved to create that sample as

$$\tau_i = \left[ \left( \sum_{l,m} \tau_{g,lm} l_{lm}^{(i)} \right) + (\lambda - \lambda_0) \left( \sum_{l,m} D_{lm} l_{lm}^{(i)} \right) \right] L. \tag{2}$$

With the proper combination of modes, we can assure a set of samples with constant incremental equivalent group delays and chromatic dispersion values, that is,

$$\tau_i = \tau_{eq,1} L + (i-1)\Delta\tau + (\lambda - \lambda_0)(D_{eq,1} + (i-1)\Delta D)L, \tag{3}$$

being $\tau_{eq,1}$ the equivalent group delay per unit length of the first sample, $D_{eq,1}$ the equivalent chromatic dispersion of the first sample, and $\Delta D$ the incremental dispersion parameter between adjacent modes.

## 3. DESIGN OF TUNEABLE TRUE TIME DELAY LINE ON A 7-MODE FIBER

We have designed a FMF for tuneable sampled TTDL operation. We choose a step-index profile to minimize the mode-coupling by increasing the effective index difference between modes with a ring-core architecture to provide more design versatility and allow us to properly manage the propagation characteristics of the symmetric modes (in a different way than the rest of the modes). Fig. 2 (a) depicts the refractive index profile of the designed FMF. The refractive index profile consists of a SiO$_2$ doped with a low-GeO$_2$ concentration inner layer (radius $a_1$ = 3 µm and inner core-to-cladding relative index difference $\Delta_1$ = 0.21%) surrounded by the SiO$_2$-GeO$_2$ doped core layer (radius $a_2$ = 10 μm and core-to-cladding relative index difference $\Delta_2$ = 0.72%) inside a pure silica cladding. The fiber supports 7 LP modes with a minimum effective index difference between them above 5·10$^{-4}$. Table 1 summarizes the main characteristics of the fiber modes at 1550-nm wavelength.

|  | LP$_{01}$ | LP$_{11}$ | LP$_{21}$ | LP$_{31}$ | LP$_{02}$ | LP$_{12}$ | LP$_{41}$ |
|---|---|---|---|---|---|---|---|
| $\tau_{lm} - \tau_{01}$ (ps/km) | 0 | 3489.08 | 8182.33 | 13022.34 | 2858.64 | 8912.83 | 17412.05 |
| $D$ (ps/km/nm) | 18.96 | 23.77 | 27.41 | 29.19 | 17.14 | 11.07 | 25.24 |
| $n_{eff}$ | 1.452726 | 1.451956 | 1.450294 | 1.448120 | 1.447556 | 1.446090 | 1.445584 |

**Table 1.** *Characteristics of the modes for the designed FMF at the wavelength of $\lambda_0$ = 1550 nm.*

The TTDL was designed to operate at a central wavelength of $\lambda_0$ = 1550 nm, in which the differential group delay per unit length was set to 100 ps/km. We use 5 mode converters based on broadband LPGs to satisfy the TTDL requirements, so the wavelength operability range of the TTDL is limited by the spectral bandwidth of the LPGs. As Fig. 2 (b) shows, the 4$^{th}$ sample travels into the LP$_{21}$ mode along the whole fiber length $L$. The rest of the samples are created by exciting the LP$_{02}$ mode at the fiber input. After a given length $L_{02} = l_{02} \cdot L$, all the power coming from this mode is coupled to LP$_{12}$ mode. The output of LP$_{12}$ mode creates the 1$^{st}$ sample, while the other samples are created by coupling part of this LP$_{12}$ mode into LP$_{01}$ and LP$_{11}$ modes by the corresponding LPGs. Two more LPGs convert these modes into the final LP$_{41}$ and LP$_{31}$ modes, whose output correspond to samples 2 and 3, respectively. With the help of Eq. (2) and following the procedure illustrated in Fig. 2 (b), we obtain the following expressions for the sample group delays

$$\begin{cases} \dfrac{\tau_1}{L} = \left(\tau_{02} l_{02} + \tau_{12} l_{12}^{(1)}\right) + (\lambda - \lambda_0)\left(D_{02} l_{02} + l_{12}^{(1)} D_{12}\right) \\ \dfrac{\tau_2}{L} = \left(\tau_{02} l_{02} + \tau_{12} l_{12}^{(2)} + \tau_{01} l_{01}^{(2)} + \tau_{41} l_{41}^{(2)}\right) + (\lambda - \lambda_0)\left(D_{02} l_{02} + D_{12} l_{12}^{(2)} + D_{01} l_{01}^{(2)} + D_{41} l_{41}^{(2)}\right) \\ \dfrac{\tau_3}{L} = \left(\tau_{02} l_{02} + \tau_{12} l_{12}^{(3)} + \tau_{11} l_{11}^{(3)} + \tau_{31} l_{31}^{(3)}\right) + (\lambda - \lambda_0)\left(D_{02} l_{02} + D_{12} l_{12}^{(3)} + D_{11} l_{11}^{(3)} + D_{31} l_{31}^{(3)}\right) \\ \dfrac{\tau_4}{L} = \tau_{21} + (\lambda - \lambda_0) D_{21} \end{cases}, \quad (4)$$

where the superindex of the normalized length $l_{02} = l_{02}^{(1)} = l_{02}^{(2)} = l_{02}^{(3)}$ has been suppressed for simplicity. Then, substituting the mode parameters of Table 1 into Eq. (4) and forcing the sample differential delays to 100 ps/km while maintaining an incremental dispersion parameter as big as possible, we obtained the position where the LPGs should be placed, as Table 2 shows. The resulting sample group delays $\tau_{eq,i} - \tau_{01}$ (normalized to LP$_{01}$ group delay) are 7882.3, 7982.3, 8082.3 and 8182.3 ps/km and the equivalent dispersions $D_{eq,i}$ are 12.10, 17.20, 22.30, 27.40 ps/km/nm, respectively for samples 1 up to 4.

| $l_{02}$ | $l_{41}^{(2)}$ | $l_{01}^{(2)}$ | $l_{12}^{(2)}$ | $l_{31}^{(3)}$ | $l_{11}^{(3)}$ | $l_{12}^{(3)}$ |
|---|---|---|---|---|---|---|
| 0.17 | 0.24 | 0.22 | 0.37 | 0.39 | 0.26 | 0.19 |

**Table 2**. *Normalized positions of the LPGs.*

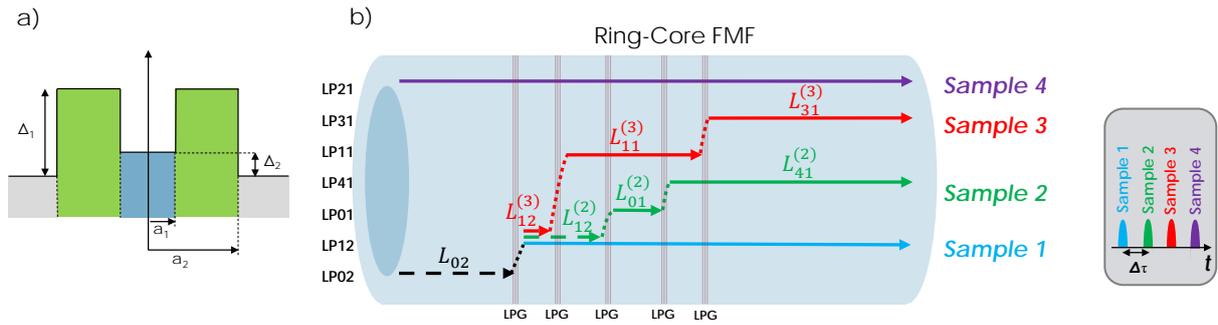

*Figure 2. a) Ring core refractive index profile. b) Scheme of the designed TTDL based on a FMF link of length L with a set of LPGs inscribed at the specific longitudinal positions and optical TTDL samples in the time domain characterized by a constant basic differential delays Δτ.*

As Fig. 3 shows, the design we have proposed, which includes a custom ring-core refractive index profile and the inscription of several broadband LPGs along the fiber, fulfils the requirements for tuneable TTDL operation along a wavelength range of 20 nm (as wide as the LPG spectral bandwidth). By changing the operation wavelength, we can tune the differential group delay between samples from 50 ps/km for 1540 nm up to 150 ps/km for 1560 nm. This tunability is obtained by controlling the slope of the spectral group delays, which corresponds to the weighted average dispersion values of the modes used. We must note that this approach design is robust against possible fabrication deviations since, once the fiber is fabricated and its propagation characteristics are known (i.e., differential group delays (DGDs) and dispersions), we can modify the LPGs positions to balance out possible DGDs and dispersions mismatches between experimental and theoretical values.

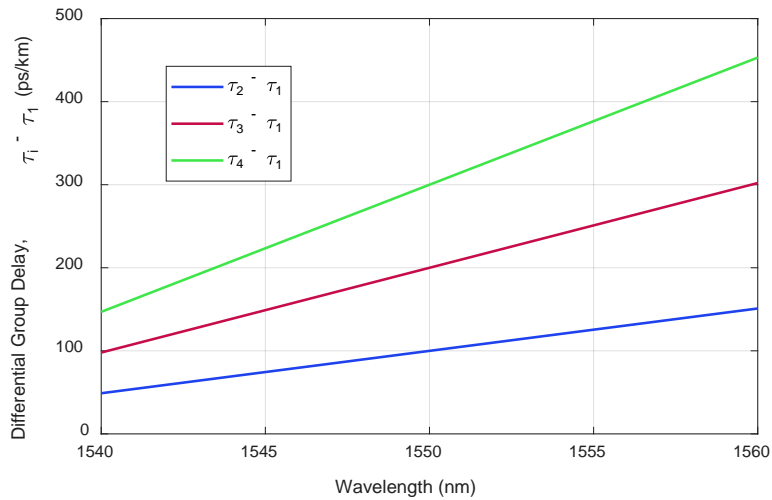

*Figure 3. Differential group delay per unit length between the TTDL samples respect to the first sample.*

## 4. CONCLUSIONS

We have presented, for the first time to our knowledge, TTDL operation for radiofrequency signals over a few-mode fiber link exhibiting time delay tuneability with the optical wavelength. The required control over both the group delay and the chromatic dispersion of the TTDL samples is achieved by the custom design of a ring-core SI FMF and the inscription of the adequate LPGs to excite higher-order modes at specific longitudinal locations along the fiber. The optical wavelength range of TTDL operation is subject to the optical bandwidth of the LPGs to be inscribed. This approach allows the implementation of tuneable MWP signal processing while the radiofrequency signal is distributed between the transmitter and the receiver ends. A particular potential field of application can be found in fiber-wireless radio access networks where different functionalities as microwave signal filtering or optical beamforming could be implemented while transmitting the signal, for instance, between a central office and a given remote antenna.

**5. ACKNOWLEDGEMENTS**

This research was supported by the ERC Consolidator Grant 724663, the Spanish MINECO Project TEC2016-80150-R, the Spanish MINECO scholarship BES-2015-073359 for S. García and the Spanish MINECO Ramon y Cajal fellowship RYC-2014-16247 for I. Gasulla.